\RequirePackage{fix-cm}
\documentclass{article}                     % onecolumn (standard format)
\usepackage{arxiv}
\usepackage[utf8]{inputenc} % allow utf-8 input
\usepackage[T1]{fontenc}
\usepackage{verbatim}
\usepackage{notoccite}
\usepackage{amsmath,graphicx}
\usepackage{makecell}
\usepackage{array}
\usepackage{multicol}
\usepackage{psfrag}
\usepackage[caption=false,font=footnotesize]{subfig}
\usepackage{multirow}
\usepackage{caption}
\usepackage[usenames, dvipsnames]{color}
\usepackage{fixltx2e}
\usepackage{picins}
\newcolumntype{x}[1]{%
>{\raggedleft\hspace{0pt}}p{#1}}%

\title{Fusion of Camera Model and Source Device Specific Forensic Methods for Improved Tamper Detection}

\author{
Ahmet~Gökhan~Poyraz \\
Dept. of Electrical-Electronic Engineering \\
Bursa Technical University, Bursa, Turkey\\
\And
A.~Emir~Dirik\thanks{Corresponding author. edirik@uludag.edu.tr}\\
Dept. of Computer Engineering\\
Bursa Uluda\u{g} University, Bursa, Turkey\\
\And
Ahmet~Karak\"{u}\c{c}\"{u}k\\
Dept. of Electrical-Electronics Engineering\\
Bursa Uluda\u{g} University, Bursa, Turkey\\
\And
Nasir~Memon\\
Dept. of Computer Science and Engineering\\
New York University, New York, USA\\
}

\begin{document}

\maketitle
\begin{abstract}
PRNU based camera recognition method is widely studied in the image forensic literature. In recent years, CNN based camera model recognition methods have been developed. These two methods also provide solutions to tamper localization problem. In this paper, we propose their combination via a Neural Network to achieve better small-scale tamper detection performance. According to the results, the fusion method performs better than underlying methods even under high JPEG compression. %Comparisons with other tamper detection schemes employing PRNU based tamper detection in the literature. 
For forgeries as small as 100$\times$100 pixel size, the proposed method outperforms the state-of-the-art, which validates the usefulness of fusion for localization of small-size image forgeries. We believe the proposed approach is feasible for any tamper-detection pipeline using the PRNU based methodology. %Because the models trained with CNN is model specific, which enable the re-use of the same CNN model between different cameras of the same model.

\end{abstract}


%removedforspringer \begin{keyword}
%removedforspringer PRNU, CNN, Neural Network, source camera, identification, source camera verification, verification, digital forensics.
%removedforspringer\end{keyword}
%removedforspringer \end{frontmatter}

\section{Introduction}
\label{sec:intro}
Digital images forgeries have become increasingly common in social media, causing a reduction of confidence in media found on the Internet. Because of this, many methods have been developed to detect tampered images, including the use of the Photo-Response Non-Uniformity (PRNU) noise based Source Camera Verification (SCV) \cite{Chen2008}.
PRNU based SCV works with a PRNU camera fingerprint which is first computed from multiple still images known to be taken by a specific camera. Then, the PRNU noise extracted from a query image is correlated with this fingerprint to determine if it was taken with the given camera. PRNU based SCV has been applied for tamper detection by matching the camera fingerprint on a block by block basis with the query image. Blocks within a tampered region of the query image will have a low correlation with the fingerprint as opposed to blocks that have not been tampered. 

Since SCV is based on a correlation operation, using PRNU for tamper detection can be challenging when the tampered region is small, especially in textured regions. There have been many approaches that have been developed to address these challenges. One such technique assumes the existence of a linear relation between image content and PRNU similarity and applies it on each image block to predict correlation values \cite{Chen2007a}. Tampering is then detected based on a significant deviation of the actual correlation from the predicted value. Another approach focuses on changes to the denoising filter and the denoiser output \cite{Chierchia2014aGuidedFiltering,Lin2016}, citing the deficiencies of the denoiser that prevent tamper detection. Some other notable attempts included refining the detector output by applying segmentation \cite{Chierchia2011}, statistical methods to improve the localization map \cite{Chierchia2014}, and multi-scale strategies \cite{Korus2016TIFSMULTISCALE} to improve localization performance.

The media forensics community has recently started to incorporate deep learning techniques, especially CNN (Convolutional Neural Network) based methods to solve media forensics problems. For example, in \cite{Liu2018}, a multi-scale CNN model was used to detect and localize forged regions. In \cite{Bondi2017}, a CNN based method for detecting forged regions from small image blocks was proposed. Also, in \cite{Bondi2017a,Tuama2016}, a method for source Camera Model Identification (CMI) using CNN was proposed and in \cite{Bayar2016} a universal manipulation detector using CNN was presented. %EMIR Say something about limitations and strength of using CNN for tamper detection. We need to setup the justification for our approach. 
However, this type of detectors require vast amounts of data for training, which may not be available in most cases and are vulnerable to adversarial attacks \cite{MARRA2018240}.

Unlike SCV which is often less than effective with small tampered regions \cite{Chen2008,Korus2016TIFSMULTISCALE}, CNN based techniques have been shown to perform well on tampered regions as small as $64\times64$ pixels \cite{Bondi2017}.%EMIR cite and indicate sizes. 
 Such CNN based techniques, as we show later, do not perform well under high compression, and PRNU based techniques do not perform well when tampered regions are with small size (e.g. $100\times100$). However, it is intuitively clear that SCV and CMI are really exploiting different aspects of an image. SCV is based on device specific characteristics that is confined to the sensor array. CMI, on the other hand, attempts to capture all the processing in the camera pipeline after the image is captured. And while the former is robust to image processing operations, the latter is effective even when presented with smaller regions of an image. In this paper, we leverage the strengths of both these approaches with the proposed method. We extensively compare the performance of SCV, CMI, and the proposed method in the assessment of the integrity of small image blocks ($96\times96$ pixels).
 
The proposed method, which we call the ``Fusion'' method, applies SCV and CMV on a block by block basis and fuses the results of the two using a neural network, to detect and localize image tampering.
The performance of the proposed method is compared with the two methods in \cite{Chierchia2014,Korus2016TIFSMULTISCALE} and is shown to yield better results with smaller forgery regions.

\section{Background}
The Fusion method incorporates traditional PRNU based SCV along with CNN based CMI, which was created from scratch albeit with inspirations from the literature.
The following sub-sections briefly introduce these methods and outline their usage within the proposed method.

\subsection{PRNU based Source Camera Verification}
\label{subsec:prnubackground}
PRNU based SCV makes use of a noise pattern called Photo-Response Non-Uniformity (PRNU) noise. The PRNU noise pattern $F$ is caused by the variations in photo-sites response to intensity of light $I_\mathrm{0}$, which occurs due to inconsistencies inherent in sensor manufacturing process. The pattern is found to be unique for each camera sensor and it is detectable from digital images produced by digital cameras \cite{Goljan2009c}. Other than PRNU, there is also random noise, denoted with $\Gamma$.

Following the simple imaging sensor output model in \cite{Goljan2009c} with matrix notation: $I=I_\mathrm{0}+I_\mathrm{0} F+\Gamma$ that incorporates both types of noises, the pattern $F$ can be estimated with a set of noise estimates $N_\mathrm{1},...,N_\mathrm{K}$, and a $denoiser$, which is commonly referred as a Wavelet denoiser \cite{Lukas2006}, s.t. $N = I-denoiser(I)$. Using these noise estimates, estimating the PRNU noise pattern of a camera's sensor $\hat F$ can be computed using the MLE estimator in \cite{Goljan2009c}. Verifying the source camera of a query image $I_q$ then only requires a Wavelet noise estimate $N_q$ of this image and the MLE-estimated PRNU pattern, $\hat F$ with normalized cross correlation $\rho=corr(N_q,I_q \hat F)$ or with peak-to-correlation-energy (PCE) formulation which applies some notable modifications \cite{Goljan2009c} to normalized cross correlation.

\subsection{CNN based Camera Model Identification}
\label{subsec:cnnbackground}
CNN consists of cascaded layers, which makes it useful to extract specific features from input images. In the convolution layer, %given or produced 
the input data has the resolution of W$_I\times$ H$_I\times$ D$_I$ (from now on, W denotes width, H height, and D depth) is convolved with a kernel with resolution W$_K\times$ H$_K\times$ D$_K$, which is initialized with random weights at the beginning and a specified value of stride. This operation produces a convolution map as the output. After the convolution layer, activation function such as Rectified Linear Unit (ReLU) is applied to avoid linearity. %Despite its simplicity, ReLU provides better performance in many CNN applications. 

In this work, the input data resolution W$_I\times$ H$_I\times$ D$_I$ is selected as 96$\times$96$\times$3, the kernel size W$_K\times$ H$_K$ is selected as 3$\times$3 and max(0,x) function is used as the ReLU activation function as shown in Fig. \ref{CNN Model}. The pooling layer carries out down-sampling operation to produce smaller feature maps. The fully connected (FC) layer carries out a dot product between the input feature vector and randomly initialized filter vector. Class scores are also produced in the FC layer. Softmax layer carries out class scores coming from FC to probability scores so that the sum of all scores becomes 1. %\vspace*{-0.4cm}

In the CNN literature, many algorithms are used for updating kernel weights. We use the most common one, the stochastic gradient descent algorithm. To train the CNN for camera model identification (CMI), we feed it with images from different camera models and labels referencing to the class of the images. %The only thing we need are images from different camera models and labels for these images for training CNN machine. 
As we want this CNN network to generate probability scores for two classes, i.e. ``target camera model'' ($\mathrm{H}_1$) and ``other camera model'' ($\mathrm{H}_0$), %However, in order to provide single camera model recognition rather than multiple camera model classifications, 
for each camera model. %The purpose here is to train a CNN machine that can detect a single camera model. 
If an image from a different camera model is given to the CNN, then it would compute a higher probability for other class, $\mathrm{H}_0$. %Therefore, the CNN based CMI method can be added combined with the outcome of the PRNU method.
%The hyper-parameters used in the CNN architecture will be explained in the next section.

\section{Proposed Method}
\label{subsec:proposedmethods}
 CNN based CMI (Method 1) basically distinguishes camera models according to the interpolation feature of the camera. Using small size blocks ($\leq$250) as input allows it to detect small-size copy-paste forgeries. PRNU-based SCV (Method 2), on the other hand, recognizes the source camera, based on the stationary features of the camera sensor. Detection of tampered regions is also possible with Method 2. However, as resolution of the selected input window decreases, performance also decreases.
We propose to combine (fuse) these two methods that can work better than each method individually. We expect that since Method 1 and Method 2 contain statistically different information, it will give a more reliable result when combined.

%trim=<left> <lower> <right> <upper>
\begin{figure}[!h]
\centering
\includegraphics[clip, trim=1.0cm 0cm 1.7cm 0.15cm, width=.7\columnwidth]{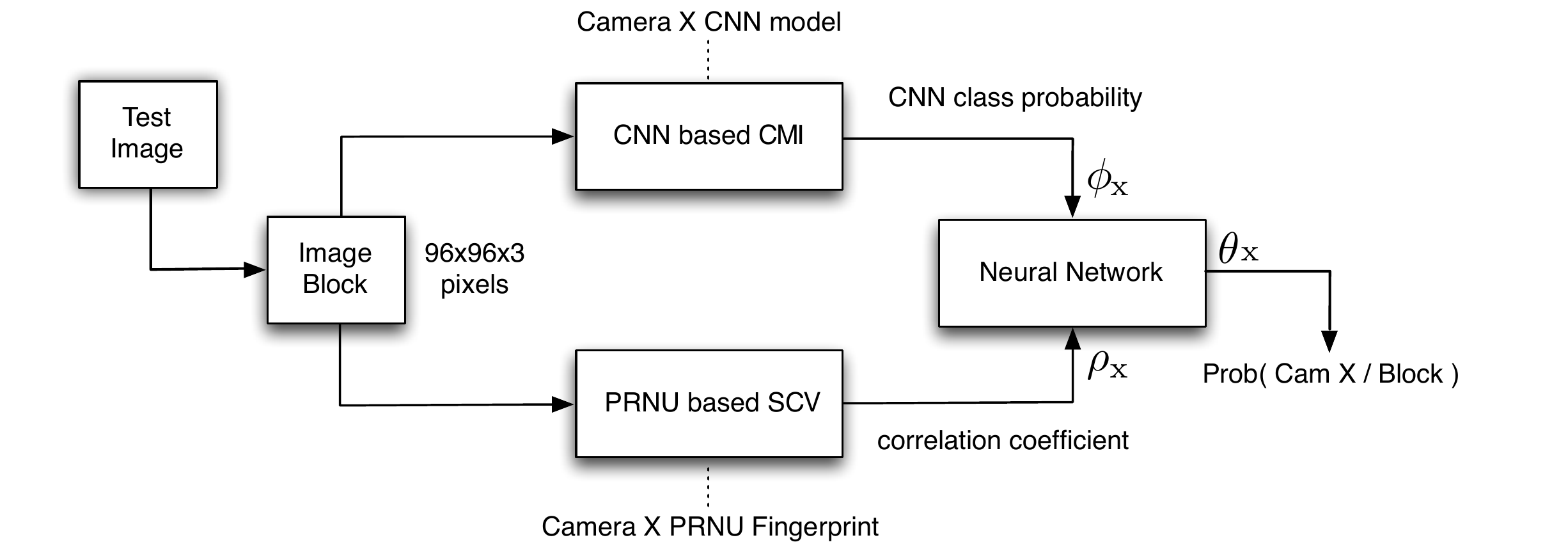}
\caption{Block Diagram for Fusion Method Approach}
\label{Fusion_Block_Diagram}
\end{figure}

Method 1 produces probability value as an output. Method 2 produces correlation between fingerprint and noise of image block. A simple NN network has been created to weight these two outputs at the best rate. Briefly, output information from Method 1 and Method 2 are given as input to NN.(Fig. \ref{Fusion_Block_Diagram})

In order to use the Fusion method, we assume that we have the model information and the PRNU fingerprint of the device. This way, we can apply both methods and produce a single outcome through the NN.

Following steps are repeated for each input image block. 

\begin{enumerate}
  \item Obtain the probability value output from CNN based CMI (Method 1), 
  \item Compute the correlation value from the PRNU based SCV (Method 2), 
  \item Combine the correlation value (from Method 1) with the probability value (from Method 2) using the proposed NN. 
\end{enumerate}

For the first step, the CNN network is trained using $96\times96$ pixel image blocks. After training, the network is able to produce probability value $\phi$ for the target camera class $H_1$  

In the second step, correlation value $\rho$, between PRNU noise in same image blocks from the first step, and their corresponding PRNU fingerprint blocks are calculated. 

In the third step, we combine the above two outcomes ($\rho$ and $\phi$). To do so, a neural network (NN) was trained which gets the values of $\rho$ and $\phi$ as input and produces a probability score for the existence of a forgery. This way, we can generate a decision map by sliding over the image $96\times96$ pixels block at a time with a small amount of shift.% to increase the resolution of the decision map. 

\begin{table}[!h]
\centering
\small\addtolength{\tabcolsep}{-4.5pt}
\begin{tabular}{lllc}
\hline
{Label} & {Make} & {Model} & {Resolution} \\ \hline
C1  & Samsung    & S3 Mini      & 2560$\times$1920 \\ \hline
C2  & Apple     & iPhone 4s     & 3264$\times$2448 \\ \hline
C3  & Huawei    & P9        & 3969$\times$2976 \\ \hline
C4  & LG      & D290       & 3264$\times$2448 \\ \hline
C5  & Apple     & iPhone 5c     & 3264$\times$2448 \\ \hline
C6  & Samsung    & Tab3       & 2048$\times$1536 \\ \hline
C7  & Apple     & iPhone 4     & 2592$\times$1936 \\ \hline
C8  & Samsung    & Galaxy S3     & 3264$\times$2448 \\ \hline
C9  & Sony-Xperia  & Z1 Compact    & 5248$\times$3936 \\ \hline
C10 & Apple     & iPad2       & 960$\times$720  \\ \hline
C11 & Huawei    & P9 Lite      & 4160$\times$3120 \\ \hline
C12 & Microsoft   & Lumia 640 LTE   & 3264$\times$1840 \\ \hline
C13 & Apple     & iPhone 6 Plus   & 3264$\times$2448 \\ \hline
C14 & Apple     & iPad Mini     & 2592$\times$1936 \\ \hline
C15 & Wiko     & Ridge 4G     & 3264$\times$2448 \\ \hline
C16 & Samsung    & Trend Plus    & 2560$\times$1920 \\ \hline
C17 & One Plus   & A3000       & 4640$\times$3480 \\ \hline
\end{tabular}
\caption{The cameras used in the experiments.}
\label{tbl:datasetandres}
\end{table}

\section{Dataset and Experimental Setup}
\label{sec:cnntraningandtest}

The Vision dataset in \cite{Shullani2017}, was preferred in this study since it has images from 29 different contemporary camera models. For each camera model, we used images labeled as "flat" and "nat" (as natural). Though there are 29 models in total, 17 camera models of 21 devices were used in this study (Table \ref{tbl:datasetandres}). Only one device was used from each camera model for training operations as a target camera. The additional four devices were used to evaluate small forgery localization performance against different devices of the same model, which is explained in Section \ref{subsec:generalization}. A total of 160 nat images were used per device in all experiments (Table \ref{tab:my-table-sets}). 100 of these nat images were used only in CNN training/testing, and up to 50 images from the remaining nat images were separated for NN training and method comparison. The rest (10 images) were used for determining the optimum F-score threshold value for comparison with the other methods in Section \ref{subsec:determinefscore}. Let us denote these 10 images as the set $\mathcal{F}$. All the image sets were mutually exclusive, in other words, no data was used more than once in any image set throughout this experiment.

\begin{table}[!h]
\centering
\begin{tabular}{llllll}
\hline
\multicolumn{2}{l}{\begin{tabular}[c]{@{}l@{}}CNN Training/\\ Testing\end{tabular}} & \multicolumn{2}{l}{\begin{tabular}[c]{@{}l@{}}NN Training/\\ Method Comp.\end{tabular}} & \multicolumn{2}{l}{\begin{tabular}[c]{@{}l@{}}Threshold\\ Decision\end{tabular}} \\ \hline
Label                                      & \#                                       & Label                                      & \#                                          & Label                                      & \#                                       \\ \hline
$\mathcal{C}_\mathrm{tr}$                                      & 80                                       & $\mathcal{S}_\mathrm{tr}$                                         & 40                                        & -                                      & -                                       \\
$\mathcal{C}_\mathrm{ts}$                                      & 20                                       & $\mathcal{S}_\mathrm{ts}$                                         & 10                                        & -                                      & -                                       \\
$\mathcal{C}$                                       & 100                                      & $\mathcal{S}$                                           & 50                                        & $\mathcal{F}$                                      & 10                                      \\ \hline
\end{tabular}
\caption{Image sets used in experiments for target camera. Here, \# denotes number of images in image sets.}
\label{tab:my-table-sets}
\end{table}

\subsection{CMI Setup (Method 1)}
For CNN training, blocks of size $96\times96$ pixels were extracted randomly from each image from two classes, s.t. 500 blocks from images within the target (correct) class, and 50 blocks from images within the other class. 
 
 \begin{figure*}[ht!]
\centering
\includegraphics[clip, trim=.1cm 19cm 10cm 5cm, width=\columnwidth]{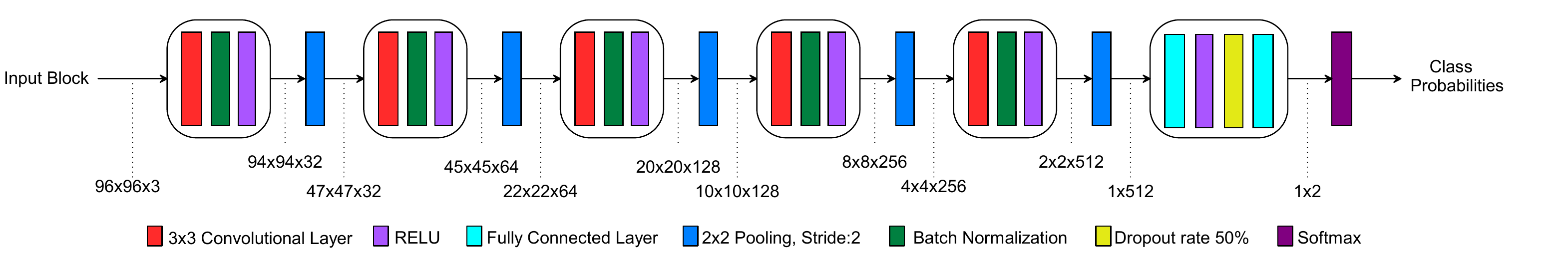}
 \caption{CNN model used in camera model classifier}
\label{CNN Model}
\end{figure*} %\vspace*{-0.8cm}

The number of blocks we used for the CNN are given in Table \ref{Training_Parameters}, where $\mathcal{C}_\mathrm{tr}$ and $\mathcal{C}_\mathrm{ts}$ denote the sets of images used for training and testing respectively, whereas the numbers show the number of image blocks in each set used. 

\begin{table}[h]
\centering
\small\addtolength{\tabcolsep}{-4.5pt}
\begin{tabular}{lllll}
\hline
                                                   & \multicolumn{2}{l}{\# of Images}                                                                                 & \multicolumn{2}{l}{\# of Blocks}                                                                                 \\ \hline
\begin{tabular}[c]{@{}l@{}}Set\\ Name\end{tabular} & \begin{tabular}[c]{@{}l@{}}Target\\ Camera\end{tabular} & \begin{tabular}[c]{@{}l@{}}Other\\ Camera\end{tabular} & \begin{tabular}[c]{@{}l@{}}Target\\ Camera\end{tabular} & \begin{tabular}[c]{@{}l@{}}Other\\ Camera\end{tabular} \\
                                                  $\mathcal{C}_\mathrm{tr}$ & 80                                                      & 1280                                                   & 40k                                                     & 64k                                                    \\
                 $\mathcal{C}_\mathrm{ts}$  & 20                                                      & 320                                                    & 10k                                                     & 16k                                                    \\
                $\mathcal{C}$  & 100                                                     & 1700                                                   & 50k                                                     & 80k                                                    \\ \hline
\end{tabular}
\caption{The data used in the CMI setup. Blocks refer to $96 \times 96$ pixels worth image patches used in CNN training and test phases}
\label{Training_Parameters}
\end{table}

In the related literature, simple models are shown to produce successful results \cite{Bondi2017CNN}. Our preliminary experiments showed CNN models with small kernel sizes perform better than larger kernels for CMI. For this reason we inspired by VGG \cite{Simonyan2014}, which employs a small size kernel filter. We experimented with different hyper-parameters and layer structures and settled with the architecture that gave the best performance, as shown in Fig. \ref{CNN Model}. The CNN network was trained for each of the 20 camera models. The training phase was stopped after 50 epochs. 

\subsection{SCV Setup (Method 2)}
For each camera model, the PRNU fingerprint was estimated using all of the available flat images. Each camera had at least 70 flat images. 
We used the PRNU method as described in \cite{Goljan2009c}. %For the PRNU fingerprint estimation, we used all of the flat images (which was 70 at the minimum) for each camera device in the vision dataset exclusively for the estimation. %We used all available flats in this dataset. Although, the number of flat images varies per camera, which at the minimum was 70 flat images. 
Please note that PRNU noise of the image blocks used here were cropped from the PRNU noise extracted from the full image.

Image blocks in the set $\mathcal{S}_\mathrm{ts}$ were used for performing PRNU based source camera identification. %$\mathcal{S}$ set is also divided into 2 parts for use in the combination part: 80\% for training the NN machine and 20\% for testing. We denote the 80\% part as $\mathcal{S}_\mathrm{tr}$ and 20\% as $\mathcal{S}_\mathrm{ts}$. 
To ensure fairness, the image blocks in the $\mathcal{S}_\mathrm{ts}$ set were used for comparison. However, as the performance of image noise extraction algorithms get worse for small image regions, we avoid extracting the noise from the image blocks in $\mathcal{S}_\mathrm{ts}$, and estimated PRNU noise $N$ from whole images. 

These noise estimates were then cropped from the identical coordinates of each $96\times96$ pixel image blocks in set $\mathcal{S}_\mathrm{ts}$. Then, they were correlated with the PRNU fingerprint blocks extracted from the identical coordinates of the corresponding camera PRNU fingerprints.

Some of the images placed in the set $\mathcal{S}_\mathrm{ts}$ were found to produce unexpectedly lower-than-threshold PRNU similarity (50 in terms of PCE) values against their matching PRNU fingerprints. We attributed such outcomes to possible mistakes in the dataset labels. For this reason, the image blocks from these images were excluded from the set $\mathcal{S}$.
%These exclusions were especially higher on the Huawei P9, Huawei P9 Lite and OnePlus A3000 devices, almost half of the images were not included for these devices. 
%However, we kept the image blocks from these cameras for comparing the methods. %It is seen here that the results obtained even with blocks extracted from less number of images are quite successful.
%eger bu adetleri vermiyorsak, yukarıda commentlenen ifadeleri etmemizin bir anlamı yok.

\begin{figure*}[ht!]
\centering
\includegraphics[width=0.75\columnwidth]{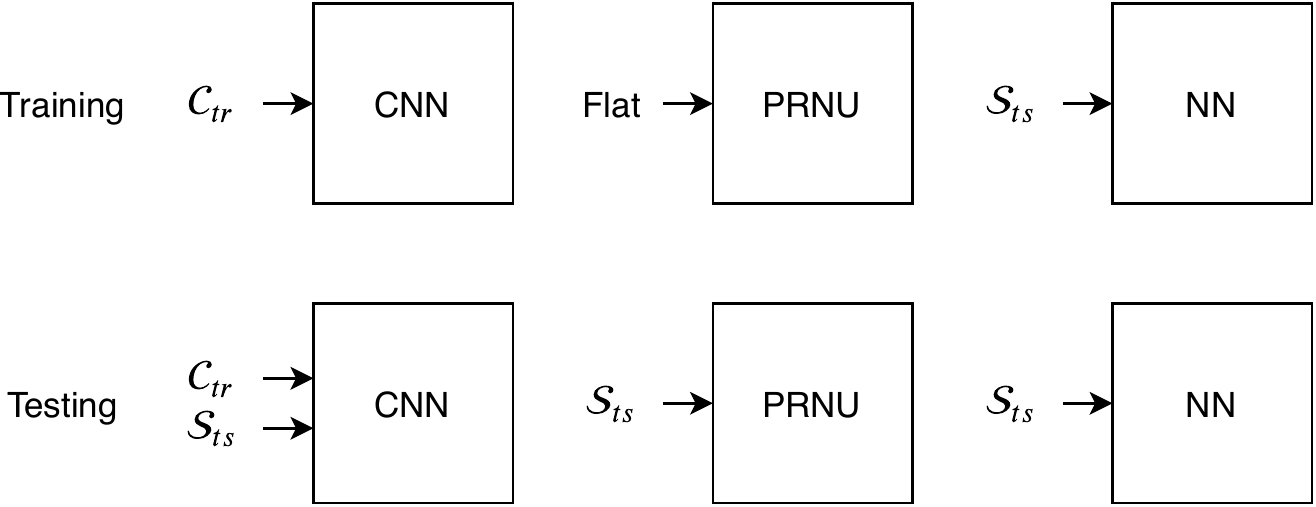}
 \caption{Illustration of image sets used during the training and testing phases.}
\label{experimenalsetup}
\end{figure*} %\vspace*{-0.8cm}

\subsection{Combining CMI and SCV outcomes}
The Fusion method comprises 3 steps in total. First, the probability value was acquired from the CNN classifier, then the correlation value from the PRNU method was computed and lastly, the NN part is trained to combine these two and produces the final probability value. Therefore, the third step forms the actual output of the Fusion method.

We construct a simple artificial NN using $\rho$ (from PRNU) and $\phi$ (from CNN classifier) values obtained from the $\mathcal{S}_\mathrm{tr}$, which includes the image blocks corresponding to $\mathrm{H_1}$ and the $\mathrm{H_0}$ cases. Therefore the NN can learn from both the matching ($\mathrm{H_1}$) and the non-matching ($\mathrm{H_0}$) cases. 

For the NN, the number of blocks we used is denoted by $\mathcal{S}_\mathrm{ts}$ and $\mathcal{S}_\mathrm{tr}$ for the number of training blocks and testing blocks as shown in Table \ref{tbl:AUC}. 

Let us denote each RGB image block in $\mathcal{S}_\mathrm{tr}$ and $\mathcal{S}_\mathrm{ts}$, by its source, i.e. the source coordinates ($\mathrm{i}$, $\mathrm{j}$) and the index of the source image ($\mathrm{k}$) with the symbol $B_\mathrm{ijk}$, the corresponding PRNU noise with $N_\mathrm{ijk}$. Similarly, the PRNU fingerprint estimate of camera $\mathrm{x}$ for the same coordinates is denoted by $F_\mathrm{ij} ^\mathrm{x} $. These are then used to generate $\rho_\mathrm{ijk}$, as expressed in the Eq. (\ref{rho_denklemi}). The CNN tests, on the other hand, generates the $\phi_\mathrm{ijk}$ values in (\ref{phi_denklemi}). Also, in the Eq. (\ref{phi_denklemi}), $\mathrm{CNNmodel^x}$ term refers to model learned from the CNN training of camera $\mathrm{x}$ and $f_\mathrm{{CNN}}$ refers to the CNN based CMI function. %\\[-7mm]

\begin{equation}\label{phi_denklemi}
\phi_\mathrm{ijk}=f_\mathrm{{CNN}}(B_\mathrm{ijk},\mathrm{CNNmodel^x})
\end{equation} %\\[-10mm]

\begin{equation}\label{rho_denklemi}
\rho_\mathrm{ijk}=corr(N_\mathrm{ijk},B_\mathrm{ijk} F_\mathrm{ij}^x)
\end{equation} 

In our experiments, we find that the model performs best with 2 hidden layers and 10 nodes for each hidden layer. $\rho$ and $\phi$ values, obtained from $\mathcal{S}_\mathrm{tr}$, were used for the NN training and the learned model will be denoted by $\mathrm{NNmodel}^\mathrm{x}$. Thus, when $B_\mathrm{ijk}$ is given as an input to $f_\mathrm{NN}$, the neural network function, the probability of tampering for a given $B_\mathrm{ijk}$, denoted by $\theta_\mathrm{ijk}$ is produced Eq. (\ref{NN_denklemi}). %\\[-7mm]

\begin{equation}\label{NN_denklemi}
\theta_\mathrm{ijk}=f_\mathrm{NN}(\rho_\mathrm{ijk},\phi_\mathrm{ijk},\mathrm{NNmodel}^\mathrm{x})
\end{equation}%\\[-7mm]
 
\section{Results and Discussion}
\label{sec:results}
In this section, we provide results of the proposed method in different settings. %In Section \ref{subsec:rocWithin} the Receiver Operating Characteristics (ROC) of PRNU based SCV, CNN based CMI and the Fusion method are compared. We further evaluate the robustness of these methods under JPEG compression in Section \ref{subsec:Robustness}. We then look at the usability of pre-trained camera models in Section \ref{subsec:generalization}. Lastly, in Section \ref{subsec:determinefscore} we 
and compare results with benchmarks using two different methods in the literature.

\subsection{Receiver Operating Characteristics of the Proposed Method}
\label{subsec:rocWithin}

\begin{table}[!t]
\centering
\small\addtolength{\tabcolsep}{-4.5pt}
\begin{tabular}{lccccc}
\hline
 & CNN & \# of & \multicolumn{3}{c}{\multirow{2}{*}{AUC Values on set $\mathcal{S}_\mathrm{ts}$}}\\
Label & Acc. & Blocks & \multicolumn{3}{c}{}\\
 & on $\mathcal{C}_\mathrm{ts}$ & $\mid\mathcal{S}_\mathrm{ts}\mid$/$\mid\mathcal{S}_\mathrm{tr}\mid$ & PRNU & CNN & Fusion\\ \hline
%\multicolumn{1}{|c|}{} & \multicolumn{1}{c|}{} & \multicolumn{1}{c|}{} & & & & \multicolumn{3}{c|}{}  \\ \cline{7-9} 
%\multicolumn{1}{|c|}{} & \multicolumn{1}{c|}{} & \multicolumn{1}{c|}{}  &  &  &        & PRNU   & CNN    & Fusion \\ \hlin
C1 &  92\%   &  13.9k / 55.6k    & 0.97  & 0.96  & 0.99 \\ \hline
C2 &  80\%   &  14.5k / 58k     & 0.95  & 0.92  & 0.99 \\ \hline
C3 &  95\%   &  10.8k / 43.2k    & 0.91  & 0.99  & 0.99 \\ \hline
C4 &  94\%   &  14.2k / 56.8k    & 0.95  & 0.98  & 0.99 \\ \hline
C5 &  93\%   &  13.7k / 54.8k    & 0.99  & 0.96  & 0.99 \\ \hline
C6 &  97\%   &  13k / 52k      & 0.98  & 0.96  & 0.99 \\ \hline
C7 &  97\%   &  14.5k / 58k     & 0.98  & 0.99  & 0.99 \\ \hline
C8 &  94\%   &  14.5k / 58k     & 0.98  & 0.98  & 0.99 \\ \hline
C9 &  99\%   &  14.5k / 58k     & 0.99  & 0.99  & 0.99 \\ \hline
C10 &  85\%   &  14.5k / 58k     & 0.94  & 0.96  & 0.98 \\ \hline
C11 &  96\%   &  10.6k / 42.4k    & 0.97  & 0.99  & 0.99 \\ \hline
C12 &  95\%   &  14.5k / 58k     & 0.96  & 0.99  & 0.99 \\ \hline
C13 &  89\%   &  14.5k / 58k     & 0.98  & 0.94  & 0.99 \\ \hline
C14 &  85\%   &  13.5k / 54k     & 0.99  & 0.91  & 0.99 \\ \hline
C15 &  92\%   &  14.5k / 58k     & 0.98  & 0.99  & 0.99 \\ \hline
C16 &  85\%   &  14.5k / 58k     & 0.95  & 0.97  & 0.99 \\ \hline
C17 &  95\%   &  12k / 48k      & 0.99  & 0.98  & 0.99 \\ \hline
\end{tabular}
\caption{The performance of underlying methods. Accuracy values are denoted with (``Acc''). The AUC values from the CNN network have been obtained from set ${C}_\mathrm{ts}$ and the AUC values were calculated from their ROC's. ${S}_\mathrm{ts}$ represents the amount of blocks in the test image set, whereas ${S}_\mathrm{tr}$ represents the amount of blocks in the training image set.}
\label{tbl:AUC}
\end{table}

In this section, we evaluate the performance of the proposed Fusion method against its underlying methods to quantify the improvement for tamper detection. To test the NN performance, the correlation value $\rho_\mathrm{ijk}$, and the class probability $\phi_\mathrm{ijk}$ values are obtained from image blocks $B_\mathrm{ijk}$ s.t. $B_\mathrm{ijk} \in \mathcal{S}_\mathrm{ts}$. %AHMET what does this mean in english?
Then the performance metrics, Area Under the Curve (AUC) and ROC for each camera model were calculated using the outputs in Eqs. (\ref{phi_denklemi}, \ref{rho_denklemi} and \ref{NN_denklemi}). These metrics were calculated by counting the number of matching blocks and non-matching blocks for various threshold values, in the range of [-1,1] for the SCV output and [0,1] for the CMI and Fusion output. %AHMET dont see hwo equation 3 is used. 
%ok

Using the same blocks, the same performance metrics were also calculated with the PRNU and the CNN methods. The AUC values are given in Table \ref{tbl:AUC}. 

%In Table \ref{tbl:ornekler}, few samples showing the performance of each underlying method are presented. Copy-Paste forgeries with size 400$\times$400 coming from different camera models were applied on these examples with to a selected image from the targeted camera model. A more detailed comparison with the other PRNU based methods will be provided later in \ref{sec:comparison}.

\subsection{Robustness under JPEG Compression}
\label{subsec:Robustness}

In our previous experiments, the Fusion method achieved high performance against small blocks from original images. However, it's not clear how well the proposed method would perform with compressed test samples. In this subsection, we provide results to evaluate the robustness of the Fusion method against compression. 

\begin{figure}[!ht]
\centering
\includegraphics[width=0.5\columnwidth]{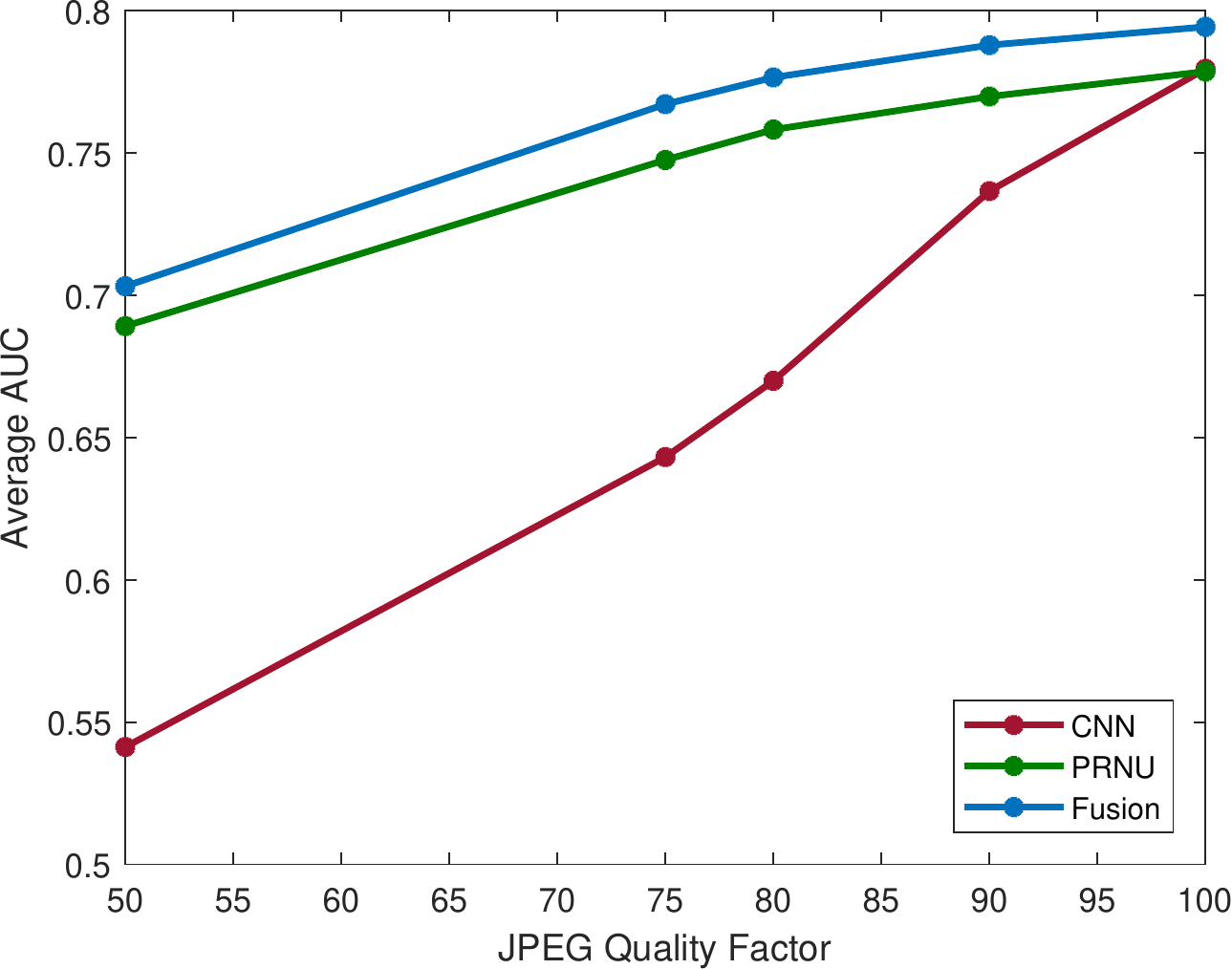}%0.48
\caption{The average AUC values for various JPEG quality factors}
\label{fig:Average_AUC}
\end{figure}

To do so, we reuse the previous experiment setup, with only one small modification: the image blocks in $\mathcal{S}_\mathrm{ts}$ were extracted from images compressed with 5 different JPEG quality factors. % as shown in Table \ref{tbl:robustness}.

It is known in the literature that the PRNU method is robust against JPEG compression. However, the same cannot be said for the CNN based CMI method as shown in Fig. \ref{fig:Average_AUC}. On images re-compressed with JPEG at various quality levels, Fusion worked reasonably well, in line with PRNU method. According to the results %in Table \ref{tbl:robustness} and 
in Fig. \ref{fig:Average_AUC}, the Fusion method performed better than both the CNN based CMI and the PRNU based SCV at all compression rates, and reached an average of 23\% improvement against CNN based CMI on images with the highest compression rate.

\subsection{Usability of pre-trained camera models}
\label{subsec:generalization}

Employing the proposed method in current tamper-detection schemes only incurs initial computation costs for unknown camera models. Thus, in other words, only new camera models (not existed in a database) need to be trained through the CNN and NN outlined in this paper. 

Recall that in the previous sections, experiments with Fusion were conducted on images strictly coming from one device for each model. In this section, we evaluate the performance of the Fusion with images from new devices with the same models where we already have NN and CNN models. % which comprised of 17 different models and 17 devices. 
%Recall that there were 21 devices in total, 
%CNN machine and the NN. Which calls for 
Although CNN models are only camera-model-specific, the NN model is trained with device-specific correlation map, which may cause it to become device-specific as well. %To do so, we performed further experiments to verify the efficiency of CMI and Fusion, to see if these pre-trained models can be re-used for different devices of the same model. 
These additional devices and their models are given in Table \ref{tbl:AUClast_section}. The test results show how the Fusion method would perform when the training device and the test device are different, but sharing the same model. The results show that the CNN based CMI and the NN models work successfully for images from another device of the same camera model. As long as we have a PRNU fingerprint of a device of a known camera model, the proposed method can be used for tamper detection regardless of the source camera device. 

Note, that the number of test images and the number of blocks were the same as in Section \ref{subsec:rocWithin}. For each device in the Table \ref{tbl:AUClast_section}, $\mathcal{S}_\mathrm{ts}$ was created and the tests were performed on this set.

\begin{table}[!ht]
\small\addtolength{\tabcolsep}{-4.5pt}
\centering
\begin{tabular}{llccc}
\hline
\multicolumn{1}{c}{Device} & \multicolumn{1}{c}{Camera} & \multicolumn{3}{c}{AUC Values on set $\mathcal{S}_\mathrm{ts}$} \\  
\multicolumn{1}{c}{Label} & \multicolumn{1}{c}{Model} & PRNU & CNN & Fusion \\ \hline
 C2,Dev-2 &  iPhone 4s  & \hspace{.1cm} 0.965 \hspace{.1cm} & \hspace{.1cm} 0.855 \hspace{.1cm} & \hspace{.1cm} 0.982 \hspace{.1cm}\\ \hline 
 C5,Dev-2 &  iPhone 5c  & \hspace{.1cm} 0.959 \hspace{.1cm} & \hspace{.1cm} 0.837 \hspace{.1cm} & \hspace{.1cm} 0.976 \hspace{.1cm}\\ \hline 
 C5,Dev-3 &  iPhone 5c  & \hspace{.1cm} 0.916 \hspace{.1cm} & \hspace{.1cm} 0.902 \hspace{.1cm} & \hspace{.1cm} 0.976 \hspace{.1cm}\\ \hline 
 C1,Dev-2 &  S3 Mini   & \hspace{.1cm} 0.962 \hspace{.1cm} & \hspace{.1cm} 0.939 \hspace{.1cm} & \hspace{.1cm} 0.984 \hspace{.1cm}\\ \hline
\end{tabular}
\caption{The details of the selected same camera models with different devices and their respective AUC values using the pre-trained models. The AUC values were calculated from ROC's. (Dev: Device)}
\label{tbl:AUClast_section}
\end{table}

\section{Tamper Detection Benchmarks}
\label{subsec:determinefscore}

In this section, we evaluate the performance of the proposed method against other PRNU based tamper detection methods in the literature.

\subsection{Methods}
For benchmarking tamper detection rates, we selected two publicly released PRNU based tamper detection methods. The first one was the multi-scale fusion (which will be denoted as ``MSF'' from now on) strategy in \cite{Korus2016TIFSMULTISCALE,Korus2016TIFSMULTISCALESTRATEGIES}. Briefly, this method produces heat-maps using the standard PRNU method through different window sizes. These heat-maps are then fused using Conditional Random Fields (CRF) thus producing binary-maps. Since each heat-map is made from windows of different sizes, smaller windows-sizes are more suitable for detecting smaller tampered regions, and similarly, larger window-sizes are better for larger tampered regions. Therefore, compared to the standard PRNU method, this method can produce better true positive and false positive rates. 

The other method used for comparison was the Bayesian-MRF method in \cite{Chierchia2014}. This method, (denoted as ``MRF'' ) replaces the constant false alarm rate with the Bayesian rule, uses BM3D non-local filtering instead of a Wavelet-based filter, and uses Markovian priors to model spatial dependencies of the image to provide better detection rate over the standard PRNU method.

Training for both methods was done using full images from the set $\mathcal{C}_\mathrm{tr}$ while CNN training of Fusion was done using blocks from the image set shown in Table \ref{Training_Parameters}. As all the three methods make use of PRNU fingerprint estimates, the same set of flat images were used to compute these estimates for each device. 

\subsection{Determining the Decision Threshold}
% tresholdlama ile akalı anlatımı proposed method altına (Sec3 ya da 4) koymamız daha isabetli olacaktır.
%To set the threshold for Fusion, 
%In earlier experiments, the probability map produced by the proposed method was used. However, as the both compared methods in this section produces binary maps.

We look for a decision threshold value to create a binary map of the output of the Fusion method for each camera model. We carry out this by measuring the F-score values, over a set of tampered images. The tampered images were generated using the image set $\mathcal{F}$, which consists of 10 images which were not used in any experiments. %AHMET this sentence needs to be improved. I cannot understand it. 

\begin{figure}[!ht]
\centering
  \subfloat[C5: iPhone 5c]{{\includegraphics[width=.45\linewidth]{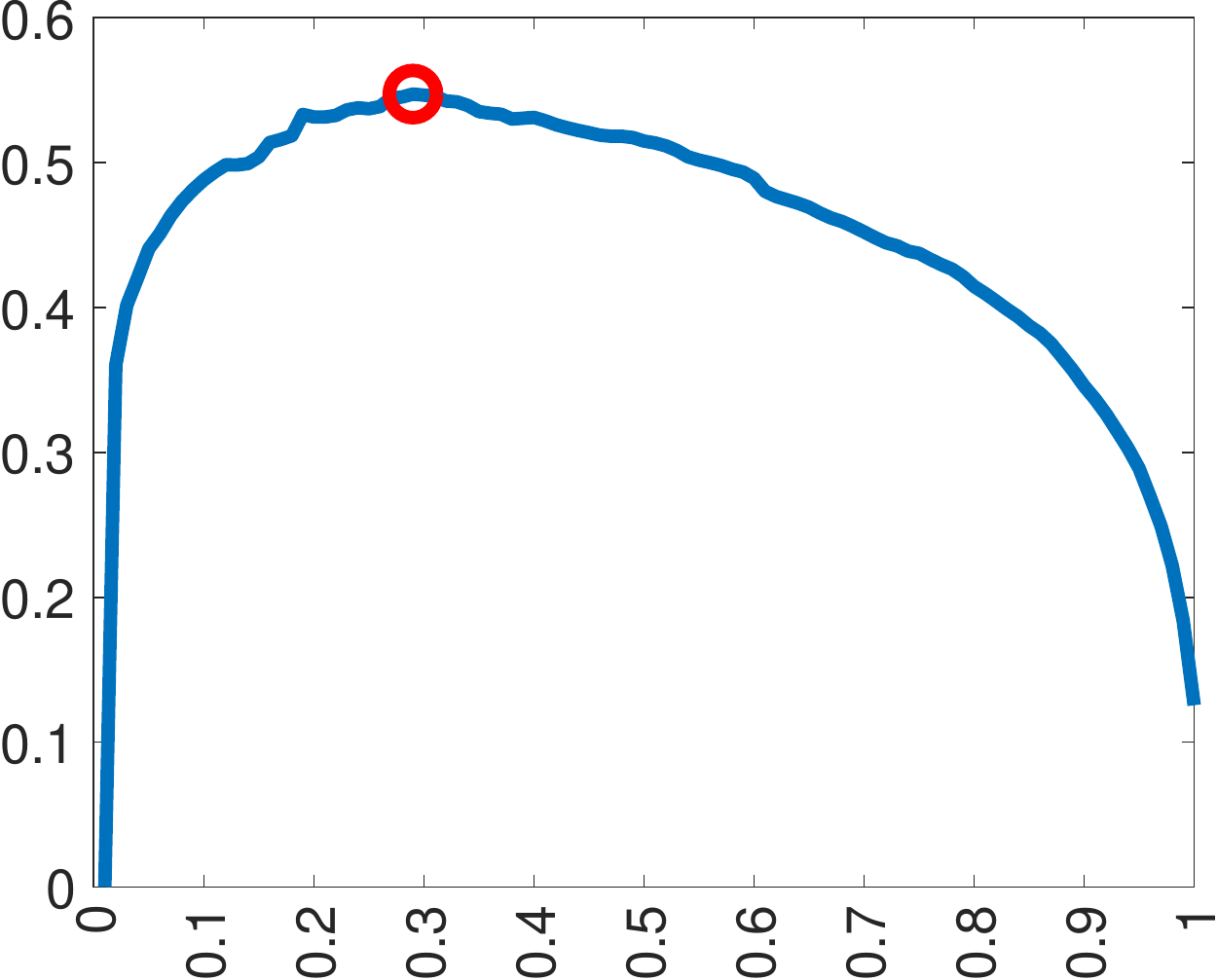} }}%
  \subfloat[C11: P9 Lite]{{\includegraphics[width=.45\linewidth]{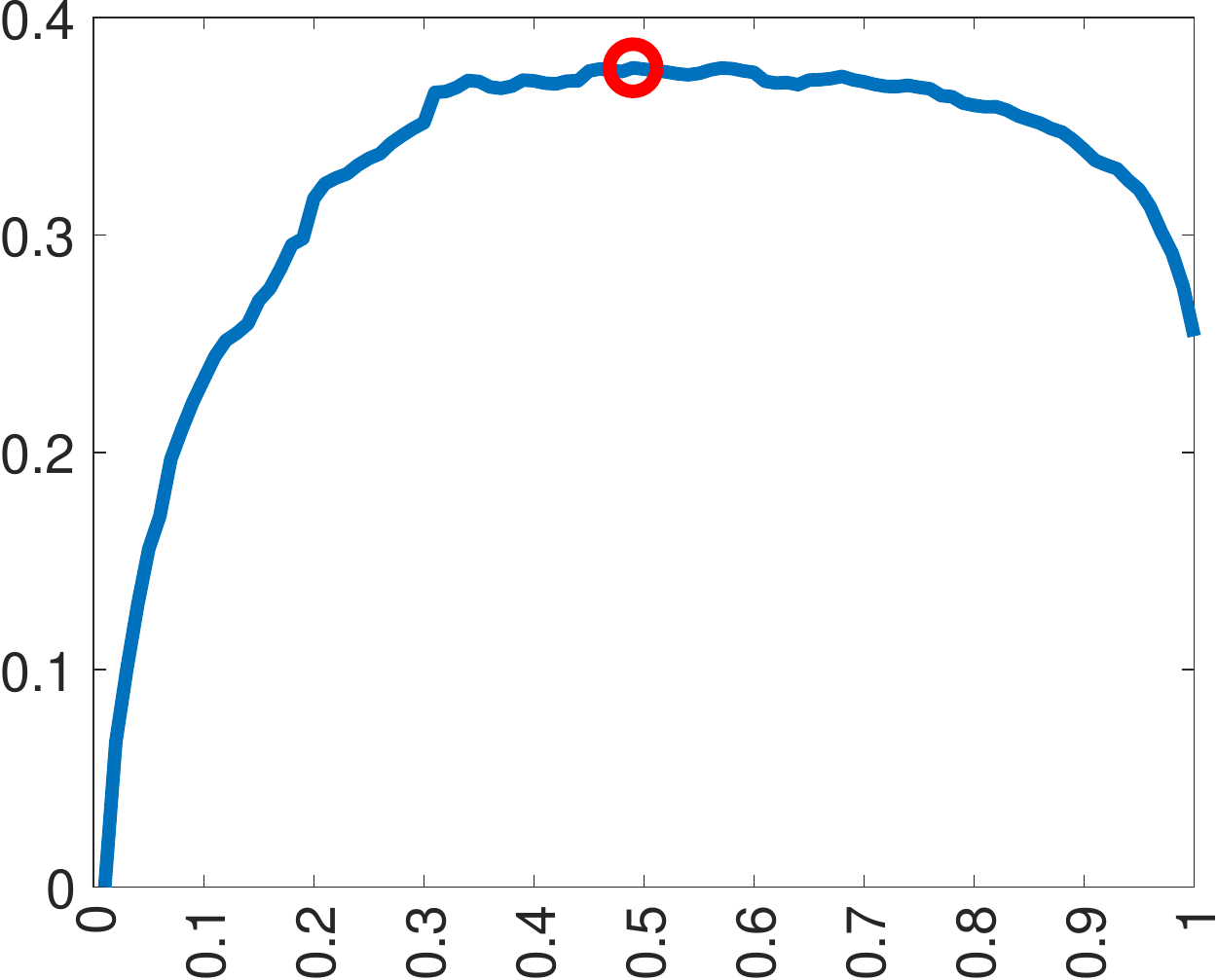} }}%
    \caption{The average F-score measurements used for determining the threshold for two example camera models. In the Figures, the threshold values which gives the maximum average F-score is selected as the decision threshold with the Fusion method and highlighted with the red marker. The horizontal axis represents the threshold values, whereas the vertical axis represents the F-score values.}
  \label{fig:fscore}
  %\centering
%\includegraphics[clip, trim=3.5cm 9cm 3.5cm 9.5cm, width=0.5\textwidth]{fskor.pdf}
\end{figure}

We applied 100 threshold values in the range of [0-1], and applied morphological opening operation through disc kernel with fixed-size (20px) radius on the binary map. Using the ground-truth for each tampered image, we then measured the F-score values. The F-score was computed with the following:
\begin{equation}
  \textrm{F-score} = \frac{2\times \textrm{TP}}{2\times \textrm{TP} + \textrm{FP} + \textrm{FN}}
  \label{eq:fscoreeq}
\end{equation}

where in Eq. \ref{eq:fscoreeq}, TP stands for true positive, FP for false positive and FN for false negative. This way, we end up having 30 different F-score measurements (3 tampered region sizes for each image is set $\mathcal{F}$) for all threshold values. The F-score measurements are then averaged and the threshold value producing the best F-score value in Eq. \ref{eq:fscoreeq} was selected as the threshold for each camera model. The averaged F-score measurements and the determined decision threshold values for two example cameras can be seen in Fig. \ref{fig:fscore}. 
%After the F-score graph of each camera device was generated, the threshold value giving the maximum f-score was applied to the comparison set.
%All comparison results are given in Table \ref{table:comparison}.

\begin{table*}[!h]
\centering
\small\addtolength{\tabcolsep}{-4.5pt}
\begin{tabular}{lccc|ccc|ccc|}
\cline{2-10}
                & \multicolumn{9}{c}{Tampered Region Size (pixels)}                                                   \\ \cline{2-10} 
                & \multicolumn{3}{|c|}{400$\times$400}     & \multicolumn{3}{c|}{200$\times$200}         & \multicolumn{3}{c|}{100$\times$100}         \\ \hline
\multicolumn{1}{|l|}{Label} & Fusion    & MRF & MSF      & Fusion    & MRF      & MSF      & Fusion    & MRF      & MSF      \\ \hline
\multicolumn{1}{|l|}{C1}   & 0.60     & 0.52 & \textbf{0.71} & 0.33     & 0.27     & \textbf{0.46} & \textbf{0.12} & 0.03     & 0.07     \\ \hline
\multicolumn{1}{|l|}{C2}   & \textbf{0.53} & 0.41 & 0.51     & \textbf{0.41} & 0.30     & 0.33     & \textbf{0.18} & 0.09     & 0.07     \\ \hline
\multicolumn{1}{|l|}{C3}   & \textbf{0.64} & 0.16 & 0.39     & \textbf{0.38} & 0.00     & 0.02     & \textbf{0.09} & 0.00     & 0.00     \\ \hline
\multicolumn{1}{|l|}{C4}   & \textbf{0.86} & 0.42 & 0.64     & \textbf{0.72} & 0.23     & 0.44     & \textbf{0.56} & 0.05     & 0.22     \\ \hline
\multicolumn{1}{|l|}{C5}   & 0.69     & 0.75 & \textbf{0.84} & 0.58     & 0.58     & \textbf{0.69} & \textbf{0.20} & 0.07     & 0.13     \\ \hline
\multicolumn{1}{|l|}{C6}   & \textbf{0.87} & 0.73 & 0.56     & \textbf{0.73} & 0.47     & 0.50     & \textbf{0.30} & 0.16     & 0.19     \\ \hline
\multicolumn{1}{|l|}{C7}   & \textbf{0.92} & 0.69 & 0.82     & \textbf{0.83} & 0.51     & 0.69     & 0.31     & 0.16     & \textbf{0.36} \\ \hline
\multicolumn{1}{|l|}{C8}  & 0.64     & 0.78 & \textbf{0.81} & 0.52     & 0.59     & \textbf{0.66} & 0.13     & \textbf{0.29} & 0.26     \\ \hline
\multicolumn{1}{|l|}{C9}  & 0.86     & 0.53 & \textbf{0.88} & \textbf{0.75} & 0.35     & 0.73     & \textbf{0.58} & 0.03     & 0.23     \\ \hline
\multicolumn{1}{|l|}{C10}  & 0.79     & 0.76 & \textbf{0.82} & 0.52     & 0.52     & \textbf{0.67} & 0.07     & 0.32     & \textbf{0.44} \\ \hline
\multicolumn{1}{|l|}{C11}  & \textbf{0.57} & 0.48 & 0.55     & 0.30     & \textbf{0.30} & 0.29     & \textbf{0.07}     & 0.00     & 0.01     \\ \hline
\multicolumn{1}{|l|}{C12}  & \textbf{0.81} & 0.49 & 0.79     & \textbf{0.62} & 0.26     & 0.53     & \textbf{0.36} & 0.05     & 0.14     \\ \hline
\multicolumn{1}{|l|}{C13}  & \textbf{0.67} & 0.49 & 0.57     & \textbf{0.45} & 0.29     & 0.43     & \textbf{0.19} & 0.15     & 0.15     \\ \hline
\multicolumn{1}{|l|}{C14}  & 0.71     & 0.51 & \textbf{0.81} & \textbf{0.38} & 0.21     & 0.37     & \textbf{0.03}     & 0.03     & 0.02     \\ \hline
\multicolumn{1}{|l|}{C15}  & \textbf{0.83} & 0.76 & 0.77     & \textbf{0.70} & 0.62     & 0.65     & \textbf{0.48} & 0.38     & 0.30     \\ \hline
\multicolumn{1}{|l|}{C16}  & 0.59     & 0.54 & \textbf{0.70} & 0.41     & 0.34     & \textbf{0.55} & \textbf{0.21} & 0.11     & 0.15     \\ \hline
\multicolumn{1}{|l|}{C17}  & \textbf{0.75} & 0.33 & 0.53     & \textbf{0.53} & 0.10     & 0.31     & 0.13     & 0.02     & \textbf{0.16} \\ \hline
\multicolumn{1}{|l|}{Avg} & \textbf{0.73} & 0.57 & 0.70     & \textbf{0.56} & 0.38     & 0.53     & \textbf{0.26} & 0.13     & 0.19     \\ \hline

\end{tabular}
\caption{The F-scores from the benchmarks, w.r.t. to tampered region sizes. Camera details were given in Table \ref{tbl:AUC}. MSF represents the Multi-Scale Fusion method \cite{Korus2016TIFSMULTISCALE,Korus2016TIFSMULTISCALESTRATEGIES}, MRF represents the Bayesian-Markovian Random Fields method in \cite{Chierchia2014}. Avg. have been calculated with F-score values from all tested images.}
\label{tbl:comparison}
\end{table*}

\begin{figure*}[!ht]
\centering
%\hspace*{-2cm}trim={<left> <lower> <right> <upper>}
\includegraphics[clip, trim=.4cm 8.4cm 5.5cm 7cm, width=1\columnwidth]{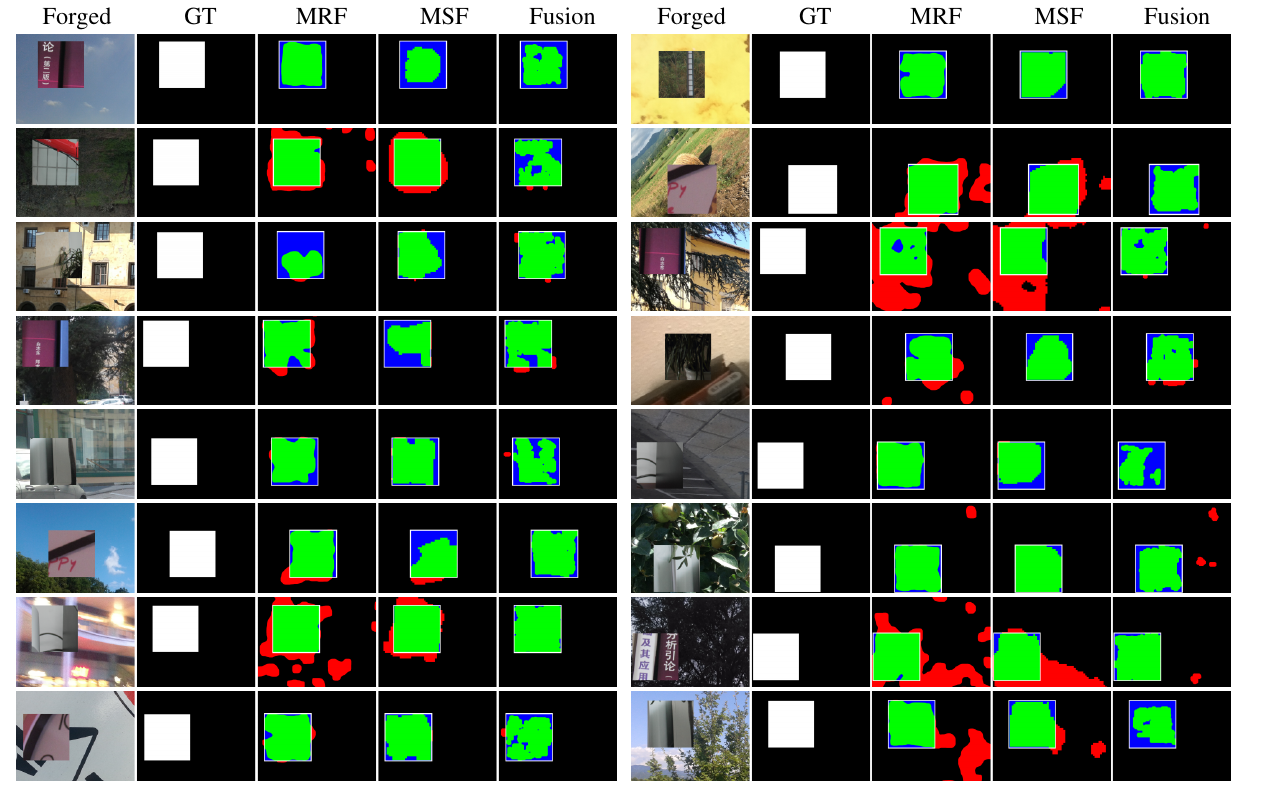}
\caption{Examples with tampered region size of 400$\times$400 pixels. In each row, there are three tampered images from one camera model, ordered from C1 to C16 from top left to bottom right. The colors on the maps are coded as the following: green - true positives; blue - false negatives; red - false positives; black - true negatives. %In column (a), the tampered region size is 400$\times$400 pixels, in (b) and in (c), the tampered region sizes are 200$\times$200 and 100$\times$100 pixels, respectively. 
Only one threshold value was picked per each camera model as explained in Section \ref{subsec:determinefscore} for Fusion method. Except for images from C10, images are cropped for better visualization.}
\label{fig:resimtablosu400}
\end{figure*}

\begin{figure*}[!ht]
\centering
%\hspace*{-2cm}trim={<left> <lower> <right> <upper>}
\includegraphics[clip, trim=.4cm 8.4cm 5.5cm 7cm, width=1\columnwidth]{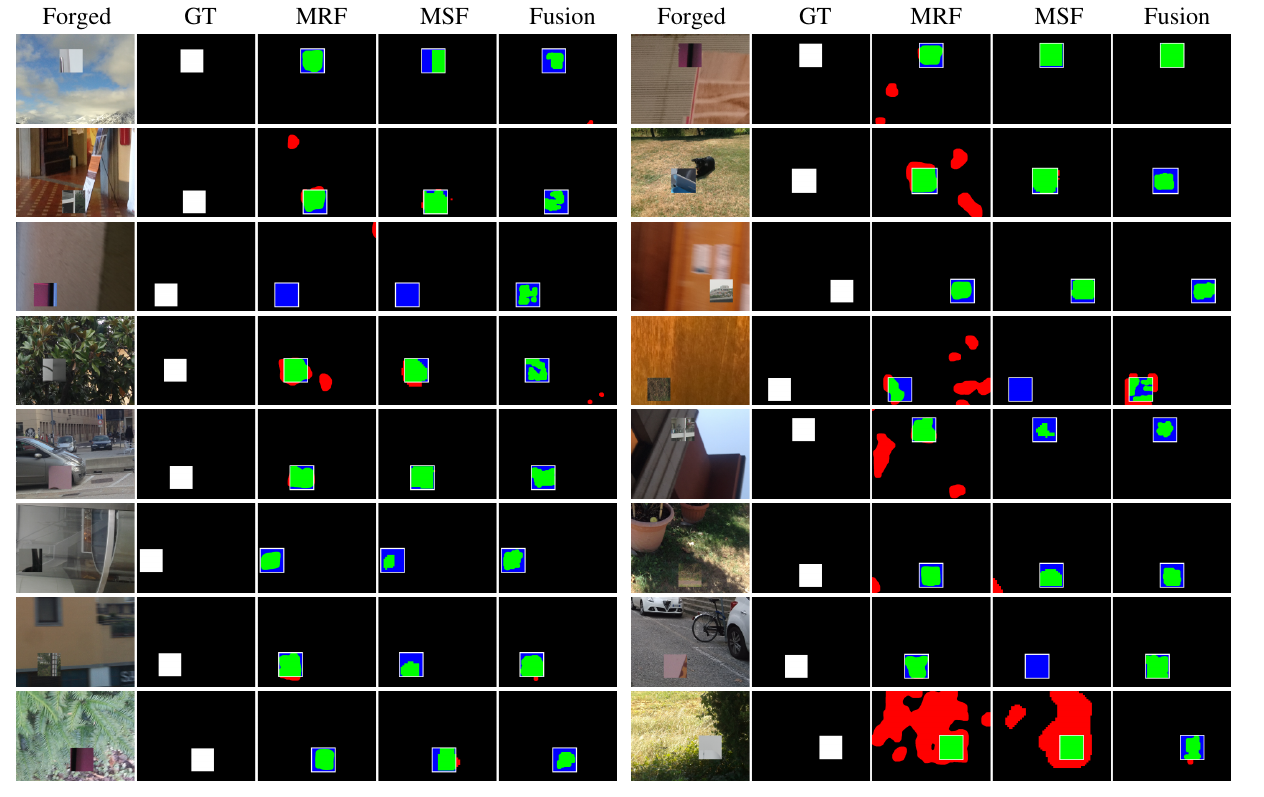}
\caption{Examples with tampered region size of 200$\times$200 pixels. In each row, there are three tampered images from one camera model, ordered from C1 to C16 from top left to bottom right. The colors on the maps are coded as the following: green - true positives; blue - false negatives; red - false positives; black - true negatives. %In column (a), the tampered region size is 400$\times$400 pixels, in (b) and in (c), the tampered region sizes are 200$\times$200 and 100$\times$100 pixels, respectively. 
Only one threshold value was picked per each camera model as explained in Section \ref{subsec:determinefscore} for Fusion method. Except for images from C10, images are cropped for better visualization.}
\label{fig:resimtablosu200}
\end{figure*}

\begin{figure*}[!ht]
\centering
%\hspace*{-2cm}trim={<left> <lower> <right> <upper>}
\includegraphics[clip, trim=.4cm 8.4cm 5.5cm 7cm, width=1\columnwidth]{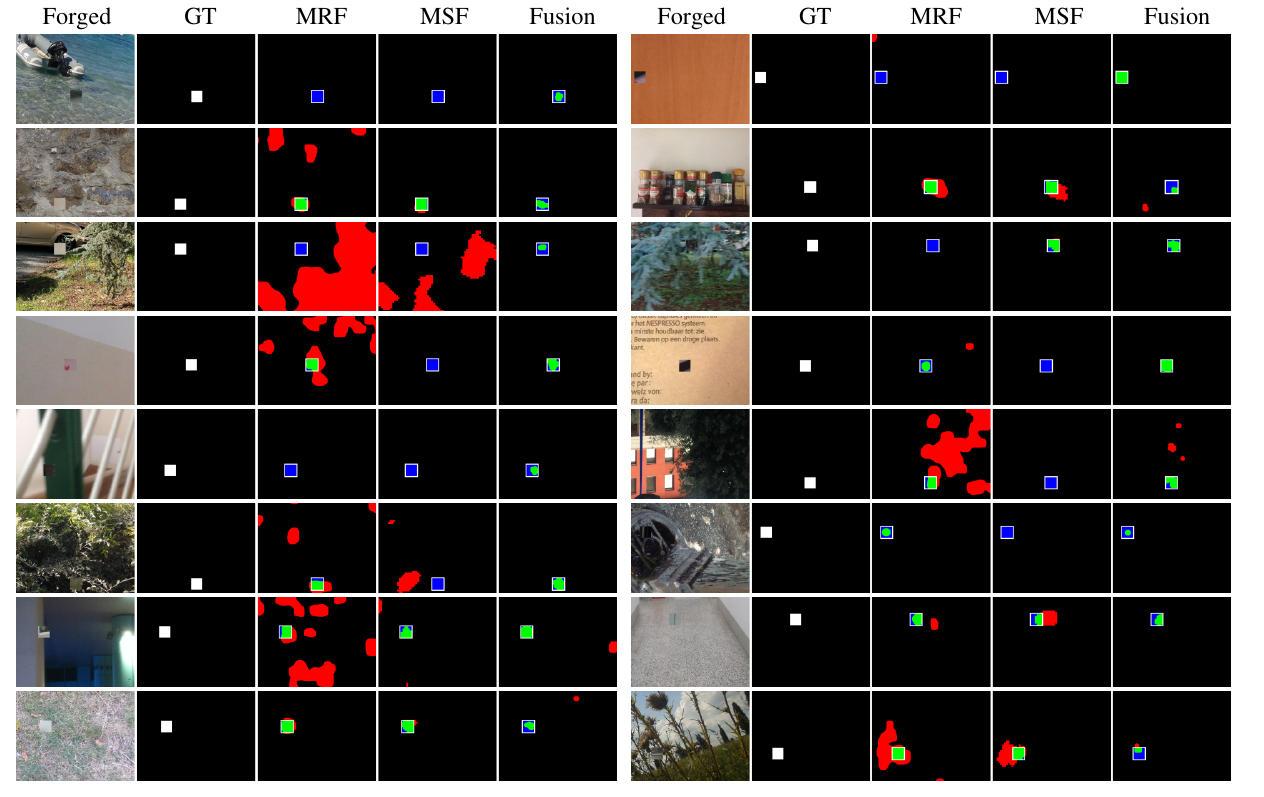}
\caption{Examples with tampered region size of 100$\times$100 pixels. In each row, there are three tampered images from one camera model, ordered from C1 to C16 from top left to bottom right. The colors on the maps are coded as the following: green - true positives; blue - false negatives; red - false positives; black - true negatives. %In column (a), the tampered region size is 400$\times$400 pixels, in (b) and in (c), the tampered region sizes are 200$\times$200 and 100$\times$100 pixels, respectively. 
Only one threshold value was picked per each camera model as explained in Section \ref{subsec:determinefscore} for Fusion method. Except for images from C10, images are cropped for better visualization.}
\label{fig:resimtablosu100}
\end{figure*}

\subsection{Comparison}
The methods were tested with the images from $S_\mathrm{ts}$ which, to recall, included 154 images across all camera devices. Note that this image set was used only for testing purposes, and no image in this set was included in any training phase of any method we compared. To evaluate tamper detection performance, we created copy-paste forgeries with the following sizes: 400$\times$400, 200$\times$200 and 100$\times$100 pixels and saved each of these images with the lossless format (PNG) leading to 462 tampered images. On average, 25 images per device were tested. Please note that for the Fusion, the same morphological treatment in Section \ref{subsec:determinefscore} was applied on threshold-applied probability maps before computing the F-score values.

In order to make the comparison fair, a window-size of 96$\times$96 pixels was chosen for the MRF method to be in line with the Fusion approach (which uses 96$\times$96). However, as the MSF method inherently utilized multiple window-sizes, for MSF, we used the following 7 different sizes: 32$\times$32, 48$\times$48, 64$\times$64, 96$\times$96, 128$\times$128, 192$\times$192, 256$\times$256 pixels as stated in \cite{Korus2016TIFSMULTISCALE}. This may seem to be advantageous for the MSF method, as it has the advantage of utilizing more heat-maps, however, we believe adhering to the literature will lead to a more objective comparison. In Fig. \ref{fig:resimtablosu100}, \ref{fig:resimtablosu200} and \ref{fig:resimtablosu400}, one tampered image for each camera model are shown for various tampered region sizes.

In Table \ref{tbl:comparison}, the average F-score values for each method are provided. As seen in the Table, The proposed Fusion method performed better for all tampered region sizes. Overall, on 51 test cases (from 17 cameras and 3 different tampered region sizes) we tested, the Fusion method was the best performing in 34 cases, whereas for MSF there were 15 such cases and for MRF there were 2 such cases. 

On experiments with the largest tampered region size setting, MSF worked best in terms of leading-cases (denoted with bold characters in Table \ref{tbl:comparison}), however, it is behind the Fusion method with 7\% in terms of F-scores. We believe the multi-scale strategy employed by the MSF increased its advantage within this setting. Furthermore, there are few exceptions where the Fusion method performed consistently worse for a few cameras: C8 (Samsung Galaxy S3) and C10 (Apple iPad).

\section{Conclusion}
\label{sec:conclusion}

In this paper, we present a new technique for forgery detection, by using both of the model-specific and the device-specific features of a camera to achieve better small-scale tamper detection performance. To reach this goal, a CNN classifier for CMI and the traditional PRNU based SCV method were combined via a Neural Network. We call this the Fusion method. 

We first study how much the proposed Fusion approach improves detection of forgeries over traditional methods and compared it against both CNN based CMI and PRNU based SCV methods. We found that Fusion performed better than both CNN based CMI (up to 10\%), and PRNU based SCV (up to 10\%) in detecting tampering using blocks originating from a different camera. 

We also show that the CNN and NN models trained for the Fusion with specific camera models can be re-used for localization of forgeries on a tampered image captured with a new device from the same camera model. Furthermore, the robustness of the Fusion approach and its underlying detectors was compared. We found that the robustness of Fusion was better than the underlying methods.% (except for 1 camera).
We believe that, training a CNN model with the knowledge of compression can increase the performance Fusion even further in such settings. 

To complete the study, the proposed Fusion approach was compared against other PRNU-based enhanced forgery detection methods in the literature. Results with the proposed method were better in locating forgeries, where, on average, Fusion had 0.50, MSF had 0.44, and MRF had 0.35 in terms of F-scores for all tampered region sizes. Similarly, Fusion performed best on the smallest tampered region size we tested (100$\times$100 pixels), where it performed 92\% better than MRF, and 39\% than MSF. However, %due to the fixed and small analysis window scale (96$\times$96), 
the performance of the proposed method w.r.t. other methods was found slightly poorer on larger tampered regions. 
The analysis window size of the proposed method can be adjusted to accommodate for larger tampered regions for such cases. %+Also, a multi-scale approach can increase the performance of the proposed method for localization of forgeries of any size. 

For future work, we will evaluate employing multi-scale strategies and compression aware deep networks to further improve the detection rates for forgeries of various size.

\section{Acknowledgments} 
This material is based on research sponsored by DARPA and Air Force Research Laboratory (AFRL) under agreement number FA8750-16-2-0173. The U.S. Government is authorized to reproduce and distribute reprints for Governmental purposes notwithstanding any copyright notation thereon. The views and conclusions contained herein are those of the authors and should not be interpreted as necessarily representing the official policies or endorsements, either expressed or implied, of DARPA and Air Force Research Laboratory (AFRL) or the U.S. Government. 

%removedforspringer     \section*{References}
%removedforspringer     \bibliography{references.bib}


\bibliographystyle{elsarticle-num}      % mathematics and physical sciences
\bibliography{references}   % name your BibTeX data base

\end{document}